\documentclass[12pt,a4paper]{article}
\usepackage{t1enc}
\usepackage[latin1]{inputenc}
\usepackage[english]{babel}
\usepackage{graphics}
\begin{document}

\title{SOLITON GENERATION AND PICOSECOND COLLAPSE IN 
SOLID-STATE LASERS WITH SEMICONDUCTOR SATURABLE ABSORBER}
\author{V. L. Kalashnikov, D. O. Krimer, I. G. Poloyko}
\maketitle
\begin{center}
International Laser Center, 65 Skorina Ave., Bldg. 17, Minsk, 220027
BELARUS, e-mail: vkal@ilc.unibel.by, http://www.geocities.com/optomaplev
\end{center}
\vspace{1pt}

\textbf{Abstract.} Based on self - consistent field theory we study a soliton generation in cw solid-state lasers with semiconductor saturable absorber. Various soliton destabilizations, i.e. the switch from femtosecond to picosecond generation (''picosecond collapse''), an automodulation regime, breakdown of soliton generation and hysteresis behavior, are predicted. It is shown that the third-order dispersion reduces the region of the soliton existence and causes the pulse oscillation and strong frequency shift.


\clearpage

\section{Introduction}

Recent advances in ultrafast solid-state lasers have resulted in sub 6-fs
pulses generated directly from the cavity of Ti: sapphire laser [1], which
is close to the theoretical limit for optical pulses. The key role to
mode-locking in this case plays the fast optical nonlinearities, such as
electronic nonlinearities, responsible, in particular, for self-phase
modulation (SPM) and self-focusing [2].

As was shown, the generation of extremely short pulses in the lasers with
self-focusing or semiconductor saturable absorber [3] is possible due to
formation of a Schr\"{o}dinger soliton. The relevant mechanism is the
balance between SPM and group velocity dispersion (GVD), which reduces the
pulse duration and stabilizes it. This mechanism termed the soliton
mode-locking [4] works together with Kerr-lensing or saturable absorption of
semiconductor thus supporting the ultrashort pulse and preventing from the
noise generation. However, the presence of dissipative laser factors, such
as saturable gain and loss, frequency filtering etc., complicates the
soliton dynamics essentially, so that various nonstationary regimes in
Kerr-lens mode-locked lasers [5, 6] and Q-switching instability in the cw
lasers with semiconductor saturable absorber [7] are possible. Up to now a
theoretical understanding of the above-mentioned issues is lacking and
challenges for investigators efforts, that can be of particular interest for
the for generation dynamics and pulse characteristics control in femtosecond
region.

Here we present the results of our study of the soliton generation in cw
solid-state lasers with semiconductor saturable absorber. The main focus of
investigation was set on the peculiarities of the transition between
femtosecond (soliton) and picosecond generation. We shall demonstrate that
such a transition is accompanied by the threshold and hysteresis phenomena.
Based on soliton perturbation theory, the numerical simulations of two
different experimental situations have been performed. The first one
corresponds to the variation of control parameters (dispersion or pump
power), when for every new value of control parameter the laser is turned on
afresh. The second situation is for continuous variation of control
parameter during a single generation session. We demonstrate also that the
third order dispersion destabilizes the soliton and produces a strong
frequency shift of generation wavelength with respect to the gain center.

\section{Model}

We analyzed the field evolution in the distributed laser system containing
saturable quasi-four level gain crystal, frequency filter, GVD and SPM
elements, two level saturable absorber and linear loss. We assumed, that in
noncoherent approximation the saturable absorber action can be described by
operator $-\frac{\gamma \exp (-E/E_{s})}{1+\partial /\partial t}$ [8], where 
$\gamma $ is the initial saturable loss, $E=\int\limits_{-\infty }^{t}\left|
a(t\prime )\right| ^{2}dt\prime \;$is the energy fluency passed through the
absorber to moment $t$, $t$ is the local time, $E_{s}$ is the loss
saturation energy fluency; the expansion of the denominator into the time
series in $\frac{\partial }{\partial t}$ is further supposed. The gain
saturation is due to the full pulse energy. The ratio of the loss saturation
energy to the gain saturation energies $\tau $ under assumption of the equal
cross sections of the generation mode at semiconductor modulator and Ti:
sapphire active medium is $1.25\times 10^{-4}$ for $E_{s}=100\;\mu J/cm^{2}$
(compare this with the figure for semiconductor saturation energy reported
in [3]).

The evolution of generation field \textit{a} obeys the following operator equation:

\vspace{1pt}

\vspace{1pt}%
\begin{eqnarray}
\frac{\partial a(z,t)}{\partial z} = \left[\alpha \exp (-\tau E)-\frac{\gamma \exp (-E)}{1+
\frac{\partial }{\partial t}}+\left[ \frac{1}{1 + \frac{\partial }{\partial t}} - 1\right]  - l\right] a(z,t)\nonumber\\
+\left[ik_{2}\frac{\partial ^{2}}{\partial t^{2}} + k_{3}\frac{\partial ^{3}}{\partial t^{3}} - 
i\beta \left| a(z,t)\right| ^{2}\right] a(z,t)
\end{eqnarray}  

were $z$ is the longitude coordinate normalized to the cavity length, or
transit number, $\alpha $ is the saturated gain, $l$ is the linear loss, $%
k_{2,3}$ are the second- and third-order dispersion coefficients,
respectively, the energy fluency is normalized to $E_{s}$ . The time is
normalized to the inverse bandwidth of the absorber $1/t_{a}$ . With this
normalization the SPM coefficient $\beta $ is $\frac{2\pi n_{2}LE_{s}}{%
\lambda nt_{a}}$, where $L$ is the length of crystal, $n$ and $n_{2}$ are
the linear and nonlinear coefficients of refractivity, respectively, $%
\lambda $ is the generation wavelength. The term in square parenthesis
stands for the frequency filtering. The transmission band of the filter and
the absorption band of semiconductor were assumed to coincide. An expansion
of Eq. (1) up to second-order in $\partial /\partial t$ results in nonlinear
Landau-Ginzburg equation [4, 8], which we do not write out here due to its
complexity.

Since an exact general solution of Eq. (1) is unknown, we sought for the
approximated quasi-soliton solution in the form $a(z,t)=a_{0}\frac{\exp %
\left[ i\omega (t-z\delta )+i\varphi z\right] }{\cosh \left[ (t-z\delta
)/t_{p}\right] ^{1+i\Psi }}$, where $a_{0}$ is the amplitude, $t_{p}$ is
the width, $\omega $ is the frequency mismatch from filter band center, $%
\Psi $ is the chirp, $\varphi $ and $\delta $ are the phase and time delays
after the full cavity round trip, respectively. In the frame of
aberrationless approach [9], the substitution of this solution in Eq. (1)
with following expansion in $t$ yields the following set of ordinary
differential equations for evolution of pulse parameters:

\vspace{1pt}

\vspace{1pt}%
\begin{eqnarray}
\frac{da_{0}}{dz} &=&a_{0}\left[ 
\begin{array}{c}
\alpha \exp \left( -\frac{\tau a_{0}^{2}}{v}\right) +\left( 1-\omega
^{2}-v^{2}\right) \gamma \exp \left( -\frac{a_{0}^{2}}{v}\right)
-l+k_{2}\Psi v^{2}-\\
\omega ^{2}-v^{2}-3k_{3}\omega v^{2}\Psi%
\end{array}%
\right] ,  \nonumber \\
\frac{d\omega }{dz} &=&\left( \Psi a_{0}^{2}\omega ^{2}+a_{0}^{2}\omega
-2v^{2}\omega -2v^{2}\omega \Psi ^{2}-\Psi a_{0}^{2}\right) \gamma \exp
\left( -\frac{a_{0}^{2}}{v}\right) -\nonumber \\
&&2v^{2}\omega \Psi ^{2}-2v^{2}\omega - 
\Psi a_{0}^{2}\tau \alpha \exp \left( -\frac{\tau a_{0}^{2}}{v}\right)
-3k_{3}v^{4}\Psi \left( \Psi ^{2}+1\right) ,  \nonumber \\
\frac{dv}{dz} &=&\frac{1}{2v^{2}}\left( a_{0}^{4}v^{2}+2v^{4}\Psi
^{2}-a_{0}^{4}+a_{0}^{4}\omega ^{2}+2a_{0}^{2}v^{2}-4a_{0}^{2}\omega
v^{2}\Psi -4v^{4}\right)\times \\
&&\gamma \exp \left( -\frac{a_{0}^{2}}{v}\right) - 
\frac{a_{0}^{4}\tau ^{2}\alpha }{2v^{2}}\exp \left( -\frac{\tau a_{0}^{2}}{%
v}\right) +3k_{2}v^{2}\Psi +v^{2}\Psi ^{2}-\nonumber\\
&&2v^{2}-9k_{3}\omega \Psi v^{2}, 
\nonumber \\
\frac{d\Psi }{dz} &=&\left( 4a_{0}^{2}\omega
v^{2}+a_{0}^{4}\Psi +4a_{0}^{2}\omega v^{2}\Psi ^{2}-2v^{4}\Psi -\Psi
a_{0}^{4}\omega ^{2}-a_{0}^{4}\omega -2v^{4}\Psi ^{3}\right)\times\nonumber\\ 
&&\frac{\gamma}{v^{2}} \exp\left( -\frac{a_{0}^{2}}{v}\right)+\frac{\Psi \alpha a_{0}^{4}\tau ^{2}}{v^{2}}\exp \left( -\frac{\tau
a_{0}^{2}}{v}\right) -2a_{0}^{2}\beta -2v^{2}\Psi -\nonumber \\
&&4k_{2}v^{2}\Psi^{2}-2v^{2}\Psi ^{3}-4k_{2}v^{2}+12k_{3}\omega v^{2}\left( \Psi ^{2}+1\right) ,\nonumber
\end{eqnarray}

where $v=1/t_{p} $ is the inverse pulse width, and the solutions for time
and phase delays:

$\delta =\left[ a_{0}^{2}\left( \omega ^{2}+v^{2}-1)-2v^{2}\omega \Psi
\right) \right] \gamma \exp \left( -\frac{a_{0}^{2}}{v}\right) -2\omega \Psi
v^{2}-2k_{2}\omega v^{2}+$

$k_{3}v^{2}\left( 5v^{2}+3\omega ^{2}-3v^{2}\Psi ^{2}\right) -a_{0}^{2}\tau
\alpha \exp \left( -\frac{\tau a_{0}^{2}}{v}\right) ,$

\vspace{1pt}

$\varphi =\left[ a_{0}^{2}\omega \left( \omega ^{2}+v^{2}-1\right)
+v^{2}\Psi \left( v^{2}-2\omega ^{2}\right) \right] \gamma \exp \left( -%
\frac{a_{0}^{2}}{v}\right) +k_{3}v^{2}\omega \times $

$\left( 2v^{2}+2\omega ^{2}-3v^{2}\Psi ^{2}\right) +k_{2}v^{2}\left(
v^{2}-\omega ^{2}\right) +\Psi v^{4}+\beta a_{0}^{2}v^{2}-2\omega ^{2}\Psi
v^{2}-$

$\omega \tau \alpha a_{0}^{2}\exp \left( -\frac{\tau a_{0}^{2}}{v}\right) .$

Assuming that pulse width is much shorter than the cavity period $T_{cav}$ ,
the equation for the gain evolution is as follows:

\vspace{1pt}

\vspace{1pt}%
\begin{equation}
\frac{d\alpha }{dz}=\left( \alpha _{m}-\alpha \right) P-\frac{2\alpha \tau
a_{0}^{2}}{v}-\frac{\alpha }{T_{g}}, 
\end{equation}

where $P=\frac{\sigma T_{cav}I_{p}}{h\upsilon }$ is the dimensionless pump
intensity, $\sigma $ is the absorption cross-section at the pump wavelength, 
$h\upsilon $ is the energy of pump photon, $\alpha _{m}$ is the maximal gain
at the full population inversion. $P=10^{-4}$ corresponds to the pump power
of 1W for $100\;\mu m$ diameter of pump mode.

The steady-state solution of Eqs. (2, 3) describes the stable ultrashort
pulses generation. The system (2, 3) was solved by the forward Euler method
with number of iterations $~10^{6} $ and the accuracy of $~10^{-6} $ .

\section{Discussion}

First, we shall study the situation when for each new value of control
parameter the generation is formed starting from the noise spike as initial
approximation.

The normalized width of the stable pulse versus GVD coefficient is presented
in Fig. 1 for different amount of SPM. It is well known that in solid-state
lasers contrary to the dye lasers a soliton mode locking with a slow
saturable absorber is impossible because of small contribution of dynamic
gain saturation (a small $\tau $ in our notations) but at the same time the
presence of SPM can provide the stable soliton mode locking [4]. As was
expected, our calculations confirmed this conclusion for the situation with
no SPM present in the system. However, with some SPM another pulse
stabilizing mechanism comes to play a role. Introducing the SPM stabilizes
the picosecond generation (region A in Fig. 1) due to the action of negative
feed-back [8] (top-left segment of curve 1 in Fig. 2). The mechanism of this
feed-back is as follows: an increased pulse intensity causes a stronger
chirping, i.e. a wider pulse spectrum and, consequently, a higher loss at
bandwidth limiting element and, reverse, a decreased intensity minimizes the
chirp and the spectral width, reducing the loss at spectral filter.

There is a section at the curves 1-4 (Fig. 1, 2) where the Schr\"{o}dinger
soliton exists within limited window of negative dispersion. The pulse
durations lie in femtosecond region with the minimum close to zero GVD, the
pulse intensities are much higher and the chirp is very small (central part
of curve 1 in Fig. 2). Calculation showed that contrary to ps-case and the
situation with non-zero third-order dispersion the soliton has no frequency
shift from filter band center and its energy is equal to critical energy for
first-order Schr\"{o}dinger soliton $E=\frac{2\left| k_{2}\right| }{%
t_{p}\beta }$ .

At some negative GVD the switch from femtosecond to picosecond regime
(''picosecond collapse'' marked by arrow, see curve 1 in Fig. 1) takes
place. Formally, these two types of ultrashort generation correspond to
domination of either nondissipative (case of Schr\"{o}dinger soliton
generation) or dissipative (case of picosecond generation) terms in Eq. 1,
respectively.

For the stronger SPM (curve 2 in Fig. 1) the interval of GVD, where the
femtosecond generation takes place, broadens which is accompanied by the
shortening of the pulse width (compare curves 1 and 2 in Figs. 1, 2). As is
seen, the ps-collapse occurs now at much higher (both in negative and
positive region) dispersions.

Further increasing of SPM transforms the character of soliton
destabilization (curve 3 in Fig. 1). In this case, the femtosecond
generation is possible only near zero dispersion; the higher negative GVD
destroys the soliton-like pulse (discontinuity of curve 3). Left to the
region where no quasi-soliton solution for Eq. 1 exists, there appears the
region of the pulse automodulational instability (region B). Here, the pulse
parameters oscillate with the period of $\sim $ $100$ cavity round-trips ($%
\sim 1$ $\mu s$ ) as is illustrated in Fig. 3. The oscillating pulse has
quasi-Schr\"{o}dinger soliton type with a small chirp (region B in Fig. 2)
and frequency shift.

Further increase of SPM (curve 4 in Fig. 1) reduces the interval of
quasi-soliton existence. Comparing curves 1 - 4 of Fig. 1 one may conclude
that the optimal amount of SPM exists providing the femtosecond generation
in the widest interval of GVD.

Let's consider the pump intensity $P\;$- another control parameter of the
system switching the generation between ps- and fs-regimes. As is seen from
Fig. 4 (triangles and arrow 1), the increase of the pump switches the laser
from ps- to fs-generation. This occurs when the pump intensity and,
consequently, the generation pulse intensity becomes high enough for SPM to
compensate the GVD.

We can investigate the system behavior depending on control parameters
(pump, GVD) by two different cases. First one is when we measure the pulse
characteristics at the some value of control parameters then turn off the
laser where upon turn on the laser and iterate this process with another
value of control parameter. Mathematically it corresponds to receiving of
the pulse as stationary solution of Eq. 1 from the noise spike. This case is
depicted at Figs. 1, 2 and triangles in Fig. 4. Second case is when one
receives ultrashort pulse at the definite value of the control parameter and
then begins smooth variation of this parameter. This case is depicted at
Fig. 4, b and Fig. 5.

As is seen from Fig. 4, the dependence of the pulse duration on the pump
intensity demonstrates a hysteresis behavior. The pump variation from the
small values to the large and backward produces the following sequence of
generation regimes (trajectory 2 in Fig. 4): the pump increase from $10^{-4}$
to $2\times 10^{-4}$ does not bring the system from ps- to fs-generation,
however the subsequent pump decrease from $2\times 10^{-4}$ to $1.67\times
10^{-4}$ switches the system to fs-regime. The range of fs-generation in
terms of $P$ (solid curve) is wider than for the case denoted by triangles,
which was plotted analogous to curves of Figs. 1 and 2. If the trajectory
starts from large $P$, the picosecond regime does not set and the
femtosecond soliton forms at $P\approx 1.67\times 10^{-4}$ (solid curve). In
this case, the interval of the soliton existence is wider than for regime
denoted by triangles.

The system demonstrates the hysteresis behavior under the variation of GVD,
too (Fig. 5): changing the GVD from the negative values to zero and than to
the positive values does not cause the switch from ps- to fs-generation. On
the other hand when moving from the positive values of dispersion to
negative and backwards one can see the dramatic change in system's behavior.
The decrease of GVD starting from the positive values, where the
ps-generation takes place, causes an abrupt switch to fs-generation near
zero GVD. Comparing curve 2 in Fig. 1 and the curve in Fig. 5 that were
plotted for the same parameters but for different physical situations
(described above), one may conclude that the region of the soliton existence
is wider in term of GVD for the latter case (when we alter the GVD) than for
the former (when generation is formed from the noise spike). The switch from
fs to ps generation under inverse movement along the GVD axis takes place at
another value of GVD thus producing a hysteresis feature (see Fig. 5).

Since the Schr\"{o}dinger soliton formation occurs at small negative GVD,
the contribution of third-order dispersion $k_{3} $ in this region may be
essential. The presence of $k_{3} \neq 0$ gives rise to the following
effects: 1) the strong pulse frequency shift which depends on the sign of $%
k_{3} $ (compare curves 2, 3 and 4 and curve 1 in Fig. 6) arises; 2) the
region of fs-generation narrows; 3) an additional destabilizing factor, the
frequency shift oscillations, appears (regions B for curves 2, 3, 4).

\section{Conclusion}

In summary, based on the self-consistent field theory we have investigated
the characteristics of Schr\"{o}dinger soliton in cw solid-state laser with
semiconductor saturable absorber. We demonstrated numerically, that the
formation of soliton has threshold character, so together with first (free
running) and second (mode-locking) threshold the threshold of femtosecond
generation exists. Three main destabilization scenarios were demonstrated:
the switch to picosecond generation (picosecond collapse), the switch to
automodulation mode and the breakdown of soliton-like pulse. The switch
between pico- and femtosecond generation has the hysteresis character. The
contribution of high-order GVD reduces the interval of quasi-soliton
existence, gives rise to the frequency shift and the pulse destabilization.

Our results may be useful for design of self-starting cw solid-state lasers
with controllable ultrashort pulse characteristics.\vspace{1pt}

This work was supported by National Foundation for basic researches (grant
F97-256).

\section{References}

1. D. H. Sutter, G. Steinmeyer, L. Gallmann, N. Matuschek, F. Morier-Genoud,
and U. Keller, ''Semiconductor saturable-absorber mirror's assisted
Kerr-lens mode-locked Ti:sapphire laser producing pulses in the two-cycle
regime'', Opt. Lett., 24, 631-633 (1999).

2. H. Haus, J. G. Fujimoto, and E. P. Ippen, ''Analytic theory of additive
pulse and Kerr lens mode locking'', IEEE J. Quant. Electr., 28, 2086-2095
(1995).

3. U. Keller, K. J. Weingarten, F. X. K\"{a}rtner, D. Kopf, B. Braun, I. D.
Jung, R. Fluck, C. H\"{o}nninger, N. Matuschek, and J. A. der Au,
''Semiconductor saturable absorbers mirrors (SESAM's) for femtosecond to
nanosecond pulse generation in solid-state lasers'', IEEE J. Selected Topics
in Quant. Electr., 2, 435-451 (1996).

4. F. X. K\"{a}rtner, I. D. Jung, and U. Keller, ''Soliton mode-locking with
saturable absorbers'', IEEE J. Selected Topics in Quant. Electr., 2, 540-555
(1996).

5. S. R. Bolton, R. A. Jenks, C. N. Elkinton, G. Sucha, ''Pulse-resolved
measurements of subharmonic oscillations in a Kerr-lens mode-locked
Ti: sapphire laser'', J. Opt. Soc. Am. B, 16, 339-344 (1999).

6. V. L. Kalashnikov, I. G. Poloyko, V. P. Mikhailov, D. von der Linde,
''Regular, quasi-periodic and chaotic behavior in cw solid-state Kerr-lens
mode-locked lasers'', J. Opt. Soc. Am. B, 14, 2691-2695 (1997).

7. C. H\"{o}nninger, R. Paschotta, F. Morier-Genoud, M. Moser, and U.
Keller, ''Q-switching stability limits of continuous-wave passive mode
locking'', J. Opt. Soc. Am. B, 16, 46-56 (1999).

8. V. L. Kalashnikov, D. O. Krimer, I. G. Poloyko, V.P. Mikhailov,
''Ultrashort pulse generation in cw solid-state laser with semiconductor
saturable absorber in the presence of the absorption linewidth
enhancement'', Optics Commun., 159, 237-242 (1999).

9. A. M. Sergeev, E. V. Vanin, F. W. Wise, ''Stability of passively
modelocked lasers with fast saturable absorbers'', Optics Commun., 140,
61-64 (1997).

\clearpage

\begin{figure}
	\begin{center}
		\includegraphics{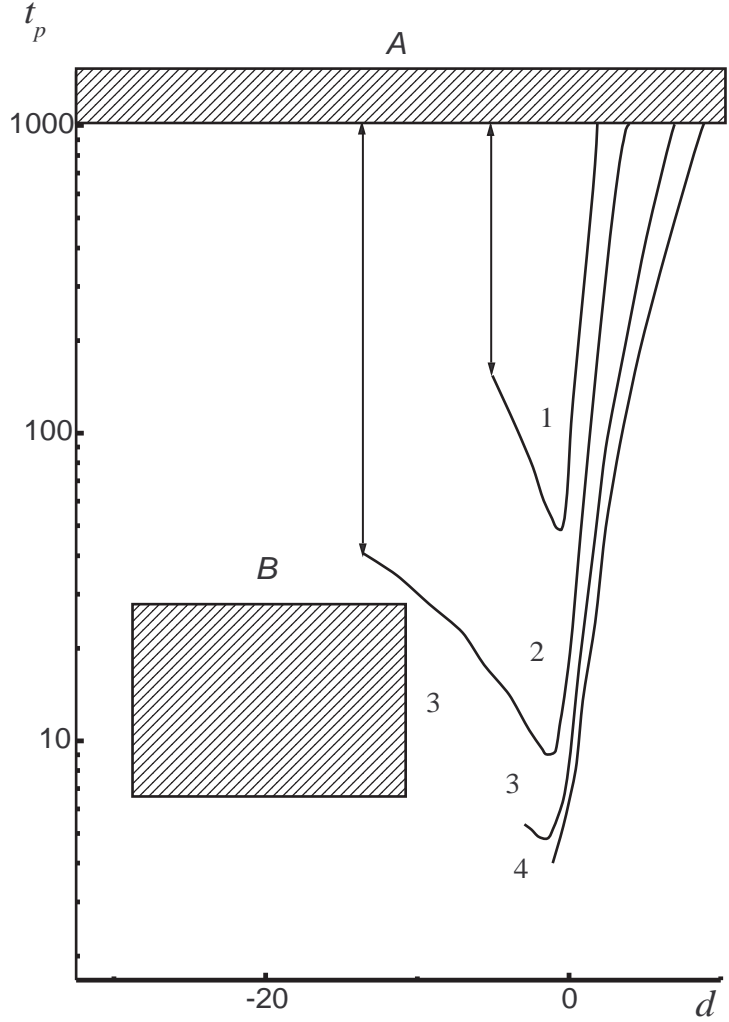}
	\end{center}

	\caption{Pulse width $t_{p}$ versus GVD coefficient $k_{2}$ (time is
normalized to $t_{a}$ ). Region A, picosecond generation, region B,
autooscillation regime for parameters of curve 3. $P=5\times 10^{-4}$ , $%
\alpha _{m}=0.5$ , $t_{a}=2.5\;fs$, $T_{g}$ $=3$ $\mu s$ , $T_{cav}=10\;ns$, 
$l=0.01$, $\gamma =0.05$ , $k_{3}=0$ . $\beta $ $=0.001$ (1), $0.01$ (2), $%
0.05$ (3), $0.1$ (4).}
\end{figure}

\begin{figure}
	\begin{center}
		\includegraphics{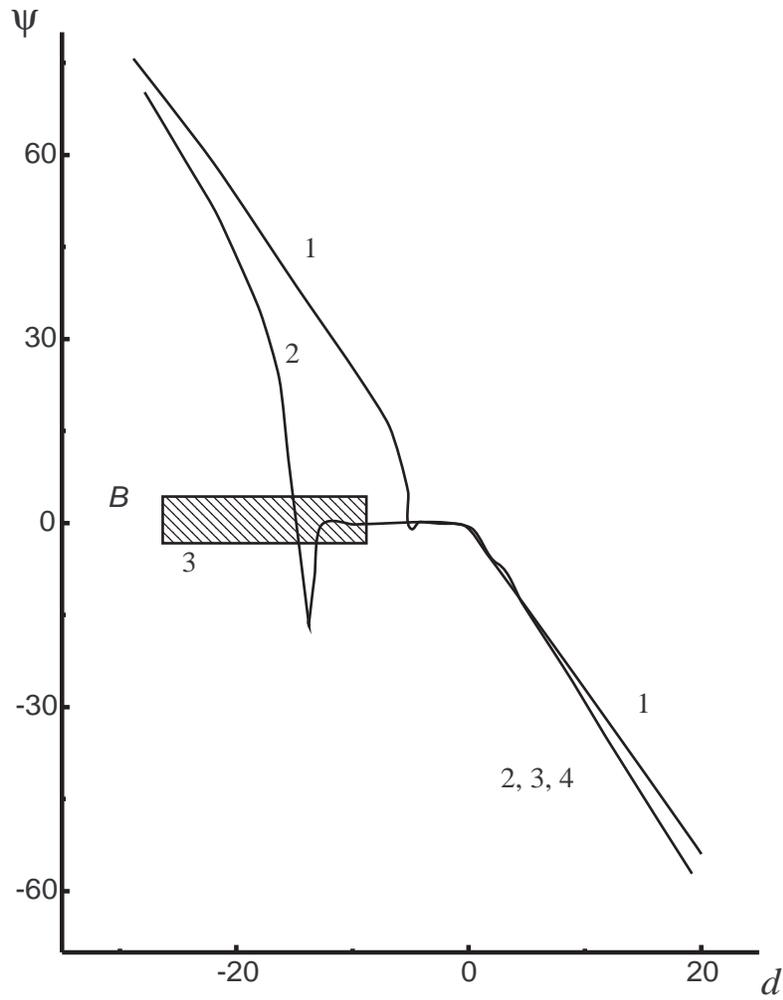}
	\end{center}

	\caption{Chirp $\Psi $ versus GVD coefficient $k_{2}$ . The parameters
correspond to Fig. 1. Curve 3 and 4 exist under $k_{2}\geq \;-3$,$\;-1$,
respectively.}
\end{figure}

\begin{figure}
	\begin{center}
		\includegraphics{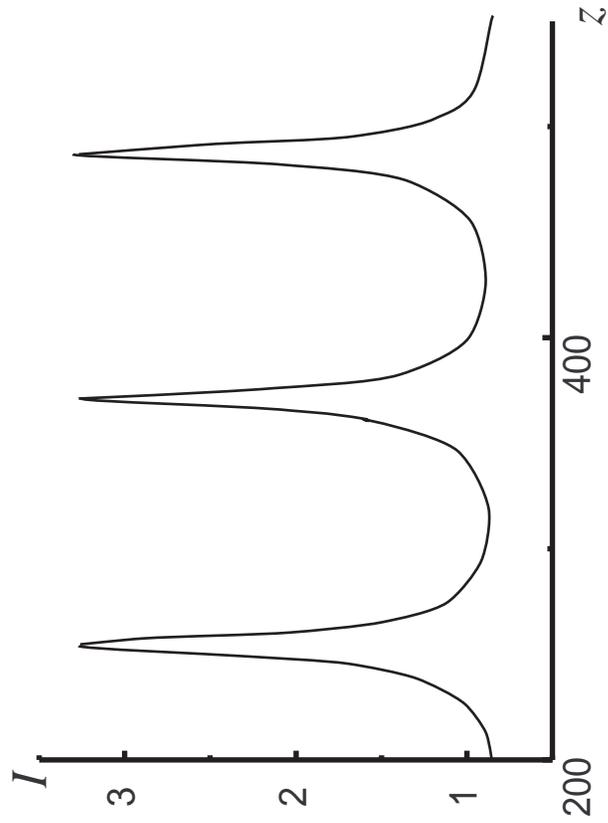}
	\end{center}

	\caption{Peak intensity $I$ normalized to $E_{s}t_{a}^{-1}$ versus round-trip number $z$. $k_{2}=-20$ , $k_{3}=0$ , $P=5\times 10^{-4}$ , $\beta $ $=0.05$.}
\end{figure}

\begin{figure}
	\begin{center}
		\includegraphics{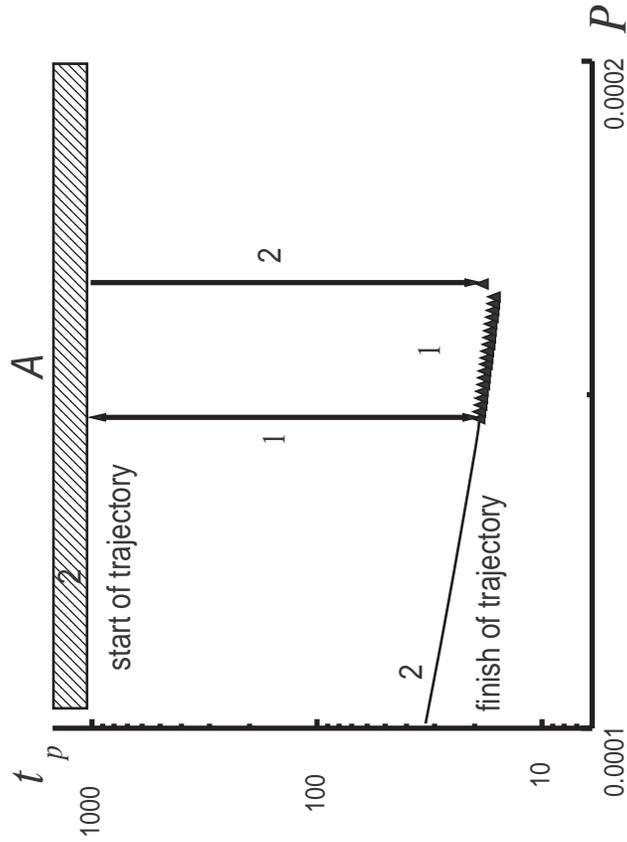}
	\end{center}

	\caption{Pulse width $t_{p}$ versus pump intensity $P$. Arrow 1 - switch
between ps- and fs-generation (triangles). $k_{2}=-10$ , $k_{3}=0$ , $\beta $ $=0.1$, other parameters correspond to Fig. 1.}
\end{figure}

\begin{figure}
	\begin{center}
		\includegraphics{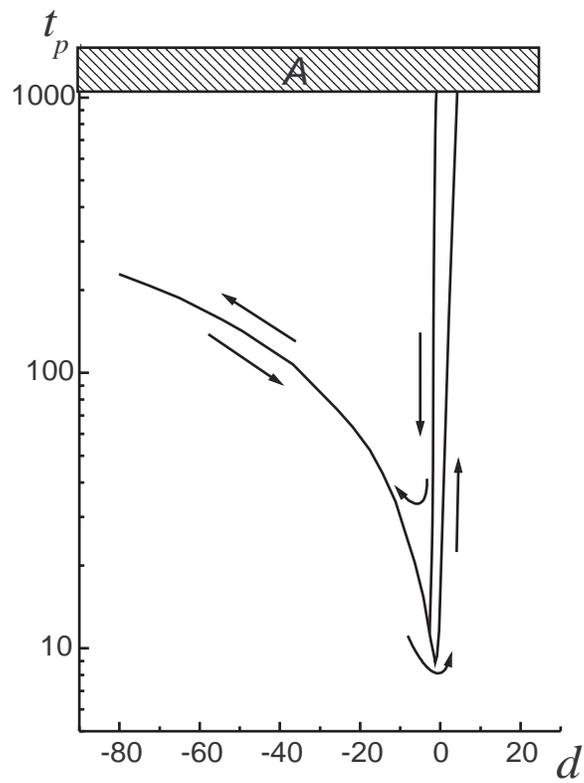}
	\end{center}

	\caption{Pulse width $t_{p} $ versus GVD coefficient $k_{2} $ . Parameters
correspond to curve 2 in Fig.1.}
\end{figure}

\begin{figure}
	\begin{center}
		\includegraphics{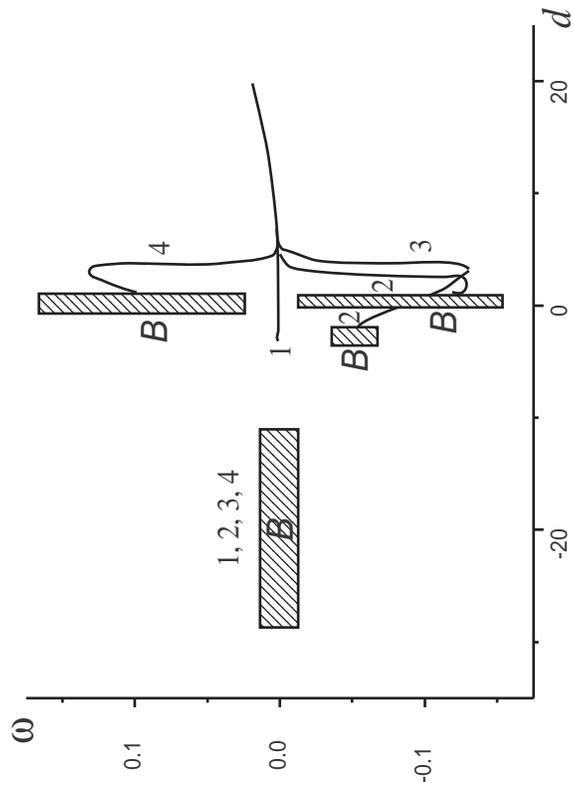}
	\end{center}

	\caption{Frequency shift $\omega $ versus GVD coefficient $k_{2}$ .
Parameters correspond to curve 3 in Fig. 1. $k_{3}=0$ (1), $-5$ (2), $-10$
(3), $10$ (4). The initial conditions correspond to Figs. 1, 2.}
\end{figure}

\end{document}